\documentclass[sigconf]{acmart}

\copyrightyear{2023}
\acmYear{2023}
\setcopyright{rightsretained}\acmConference[IUI '23]{28th International Conference on Intelligent User Interfaces}{March 27--31, 2023}{Sydney, NSW, Australia}
\acmBooktitle{28th International Conference on Intelligent User Interfaces (IUI'23), March 27--31, 2023, Sydney, NSW, Australia} 
\acmDOI{10.1145/3581641.3584036} 
\acmISBN{979-8-4007-0106-1/23/03}

\usepackage{algorithm} 
\usepackage{algpseudocode}
\usepackage{enumitem}
\usepackage{xspace}
\usepackage{wrapfig}
\AtBeginEnvironment{quote}{\itshape}

\makeatletter
\newcommand{\labeltext}[3][]{%
    \@bsphack%
    \csname phantomsection\endcsname
    \def\tst{#1}%
    \def\labelmarkup{\emph}
    \def\refmarkup{}%
    \ifx\tst\empty\def\@currentlabel{\refmarkup{#2}}{\label{#3}}%
    \else\def\@currentlabel{\refmarkup{#1}}{\label{#3}}\fi%
    \@esphack%
    \labelmarkup{#2}
}
\makeatother

\newcommand{\nasa}[0]{NASA\xspace}

\newcommand{\pixl}[0]{PIXL\xspace}

\newcommand{\pixlise}[0]{PIXLISE\xspace}

\graphicspath{ {./Images/} }

\AtBeginDocument{%
  \providecommand\BibTeX{{%
    \normalfont B\kern-0.5em{\scshape i\kern-0.25em b}\kern-0.8em\TeX}}}

\begin{document}

\title{Lessons from the Development of an Anomaly Detection Interface on the Mars Perseverance Rover using the ISHMAP Framework}

\renewcommand{\shorttitle}{Lessons from Anomaly Detection on the Mars Rover using ISHMAP}


\newcommand{\authorgap}{\hspace{3pt}}


\author{Austin P. Wright}
\affiliation{
  \institution{Georgia Institute of Technology}
  \city{Atlanta}
  \country{United States}
}

\author{Peter Nemere}
\affiliation{
    \institution{Queensland University of Technology}
  \city{Brisbane}
  \country{Australia}
}
  
\author{Adrian Galvin}
\affiliation{
\institution{Jet Propulsion Laboratory}
  \city{Pasadena}
  \country{United States}
}
\author{Duen Horng Chau}
\affiliation{
  \institution{Georgia Institute of Technology}
  \city{Atlanta}
  \country{United States}
}

\author{Scott Davidoff}
\affiliation{
\institution{Jet Propulsion Laboratory}
  \city{Pasadena}
  \country{United States}
}

\renewcommand{\shortauthors}{Wright, et al.}

\begin{abstract}
    While anomaly detection stands among the most important and valuable problems across many scientific domains, anomaly detection research often focuses on AI methods that can lack the nuance and interpretability so critical to conducting scientific inquiry. We believe this exclusive focus on algorithms with a fixed framing ultimately blocks scientists from adopting even high-accuracy anomaly detection models in many scientific use cases. In this application paper we present the results of utilizing an alternative approach that situates the mathematical framing of machine learning based anomaly detection within a participatory design framework. In a collaboration with \nasa scientists working with the \pixl instrument studying Martian planetary geochemistry as a part of the search for extra-terrestrial life; 
we report on over 18 months of in-context user research and co-design to define the key problems \nasa scientists face when looking to detect and interpret spectral anomalies. We address these problems and develop a novel spectral anomaly detection toolkit for \pixl scientists that is highly accurate (93.4\% test accuracy on detecting diffraction anomalies), while maintaining strong transparency to scientific interpretation. We also describe outcomes from a yearlong field deployment of the algorithm and associated interface, now used daily as a core component of the \pixl science team's workflow, and directly situate the algorithm as a key contributor to discoveries around the potential habitability of Mars. 
Finally we introduce a new design framework which we developed through the course of this collaboration for co-creating anomaly detection algorithms: Iterative Semantic Heuristic Modeling of Anomalous Phenomena (ISHMAP), which provides a process for scientists and researchers to produce natively interpretable anomaly detection models. 
 This work showcases an example of successfully bridging methodologies from AI and HCI within a scientific domain, and provides a resource in ISHMAP which may be used by other researchers and practitioners looking to partner with other scientific teams to achieve better science through more effective and interpretable anomaly detection tools.

\end{abstract}




\maketitle
\section{Introduction}

Detecting and interpreting anomalies is one of the core tasks at the heart of the scientific process\cite{kuhn1970structure}. However, even while novel machine learning models are rapidly improving the state of the art in anomaly detection\cite{deepAnomalySurvey}, scientific applications present an interesting and complex problem for anomaly detection. Because the goal of science is often not just to detect anomalies, but to understand their causes, developing both high-accuracy and interpretable anomaly detection models becomes one of the key ways to support scientific inquiry. While research has explored how algorithms can be made to be explain their reasoning\cite{gunning2019xai,lundberg2017unified, ribeiro2016should}, this paper investigates how HCI methods can be deeply integrated within the framing of model development to drive  interpretability as a user-defined quality that is considered a first class objective rather than a post-hoc computed explanation.  

In this work we present an application of just such a design process, we engaged in a multi-year collaboration with a team of scientists at \nasa
who analyze data from the \pixl  instrument to understand Martian geochemistry, and thus the possibility of extra-terrestrial life \cite{astrobioSurvey2005, fairen2010astrobiology, banfield2001:biosignatures, bishop2004multiple}.
Our collaboration set forth with the following research questions: 
\begin{enumerate}
    \item \textbf{RQ1:} Within the specific scientific workflow of \pixl scientists, what are the particular requirements that a modeling approach must satisfy?
    \item \textbf{RQ2:} In what ways and to what degree does the existing standard approach to anomaly detection through machine learning satisfy and violate these requirements?
    \item \textbf{RQ3:} How might a different anomaly detection modeling framework to enable the development of more effective systems given these requirements?
\end{enumerate}

In the course of addressing these questions we explored broader approaches to anomaly detection in a scientific context and report the following four contributions:

\begin{enumerate}

\item\textbf{Formative design study}. 
We present the findings of an 18 month long study where through a series of contextual inquiry interviews we outlined the analytic workflow of the \nasa \pixl science team, found key challenges faced by scientists in detecting and interpreting spectral anomalies in X-ray florescence (XRF) data, and  developed a comprehensive model of how \pixl scientists approach anomaly detection. This study revealed three key design goals used to guide the development of tools assisting in this workflow.

\item\textbf{Novel spectroscopy anomaly detection algorithm}. We describe a new method to automatically detect and classify diffraction and other spectral anomalies accurately (93.4\% test accuracy), directly within raw unquantified spectra, providing a significant improvement over existing methods.

\item\textbf{Deployed algorithm and visualization system }. We embedded this algorithm within a visualization tool that has been deployed and has now become a regular and important component of all \pixl based analysis conducted over the past year,
used daily by over 97 \nasa scientists and \nasa-affiliated scientists around the globe. We evaluate the success of the application further by examining some examples of novel planetary science enabled by the tool.

\item\textbf{We introduce a novel framework, ISHMAP, for the collaborative development of anomaly detection tools for scientific teams like \pixl}. 
Finally we present a framework, Iterated Semantic Heuristic Modeling of Anomalous Phenomena (ISHMAP), which was developed from the success of this application that can serve as a useful tool in itself for future collaborations in similar scientific anomaly detection settings. This framework integrates HCI and AI perspectives on model development and presents a different formulation of the problem of anomaly detection specifically designed to fulfill the needs of scientific users. ISHMAP introduces a process for how to produce anomaly detection models that provide first-class scientific interpretations by default, ensuring scientific users buy-in, and also can be tightly integrated with existing modeling techniques (including deep learning approaches).

\end{enumerate}

This in diving deep into this specific application, this paper looks showcase an example of a possible way to synthesize AI anomaly detection research with methods developed in HCI. We believe that by combining methods across disciplines, researchers may be better able to take on high priority problems like anomaly detection, in partnership with scientific communities, and to help drive discovery. We offer evidence of how this bridging supported our own inquiry in the case of the \nasa \pixl science team. In this next section, we situate this work within the larger landscape of HCI, AI, and Astrobiology research.

\begin{figure}
  \begin{center}
    \includegraphics[width=1\columnwidth]{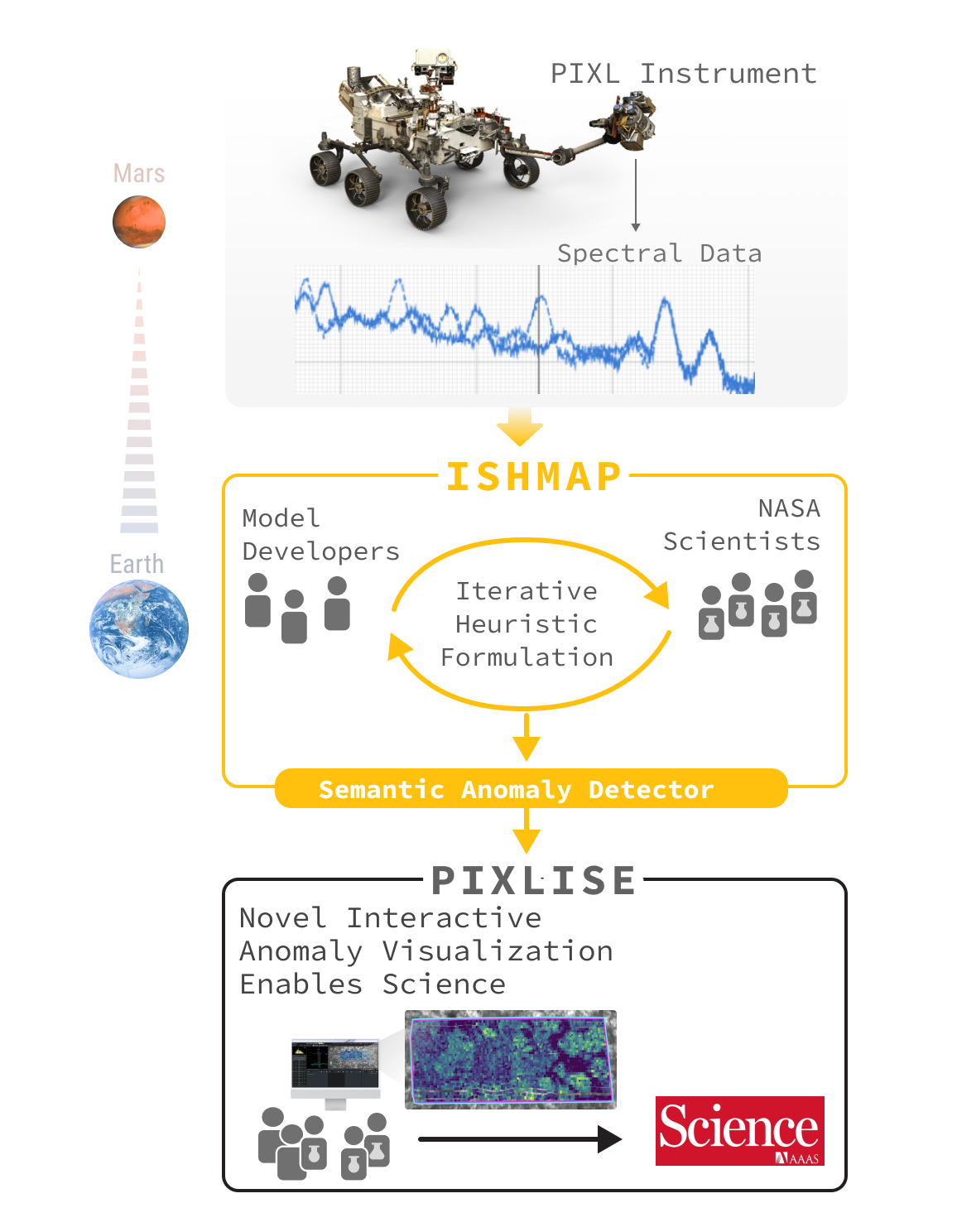}
  \end{center}
  \caption{Overview of how we utilized ISHMAP to assist in the \pixl science mission. Using this collaborative process we were able to develop a novel interpretable anomaly detection model and deploy interactive visualizations within the widely used \pixlise visual analytics program. This deployment proved to provide key insights in ongoing major scientific findings.}
  \label{fig:overview}
\end{figure}

\section{Background and Related Work}

\subsection{The Search for Extraterrestrial Life}

The search for extraterrestrial life is among the great contemporary scientific endeavors \cite{astrobioSurvey2005}, and Mars forms a key component of that search \cite{fairen2010astrobiology}. 
A principal way scientists around the world build an understanding of Mars as a possible host for life is to study its planetary-scale geology and geochemistry over time \cite{banfield2001:biosignatures,bishop2004multiple,mcmahon2018marsfieldguide}. 
\nasa's \pixl instrument supports that ongoing investigation by capturing co-aligned visible imaging and thousands of spatially localized pairs of \textit{X-ray fluorescence} (XRF) spectra in a single experiment \cite{Allwood2020}.

The \pixl instrument represents a generational change in the scale and sensitivity of extra-terrestrial XRF measurements\cite{Rieder1771,gellert2006alpha,VanBommel2016}. While this can bring analytical leaps, it also means that the data are more sensitive to spectral anomalies that can lead scientists to misinterpret data.

Anomalies in XRF spectra have been historically identified manually. But with each experiment site including thousands of spectra to manually investigate, each with hundreds of peaks, anomalies become increasingly difficult and time consuming to manually identify. And missing spectral anomalies could lead to the misinterpretation of the elemental chemistry, mineralogy and ultimately the planetary scale environment that acted upon the samples under investigation. 
Therefore there is substantial scientific interest in performing reliable and interpretable anomaly detection on the incoming data from \pixl.

\subsection{Bridging HCI and AI methods for Interpretable Modeling}

While having more independent origins, the recent history of hybrid disciplines such as HCI-AI\cite{inkpen2019aihci}, HCML \cite{gilles2016hcml}, and human-guided ML \cite{gil2019towards} reflect an interest in drawing on knowledge generated across fields to jointly inform the development of systems that touch on each discipline. HCI researchers, for example have looked to understand the challenges to design AI systems that fit with user needs\cite{yang2020reexamining}, and to use new properties exposed by these systems as a resource for designers\cite{benjamin2021mlmaterial}. The complexity of dealing with some form of embedded intelligence led other researchers to introduce particular methods to structure ideation and iteration of AI- and ML-systems for ubicomp\cite{davidoff2007speeddating,odom2012fieldwork} or dialog systems\cite{yang2019sketching}. 

Alternatively, researchers in AI and ML have drawn upon expert knowledge to inform ML models \cite{imlDataExfiltration2020}, or to bring interactivity into learning systems \cite{interactiveMachineLearning2003}, looking to leverage interaction to define ML model rules \cite{guo2022building}. These methods have been applied to anomaly detection in cybersecurity, spatio-temporal, and behaviour modeling contexts\cite{jiang2019recent}.

In the sciences, this crossover has looked at interactivity and glass-box models as a way to support interpretable and configurable deployed machine learning models \cite{wang2022interpretability}. However, fieldwork in disciplines such as oceanography have shown that while interpretation and understanding of code and models is essential, it is insufficient to contributing within a larger scientific workflow as the primary driver of change will often come from anomalous ``moments of flux'' which naturally lead to reconceptualisations that are not amenable to fixed data perspective implicit in any traditional data science or machine learning model, as ``a singular focus on problem-solving may marginalize opportunities for innovation that
could drive community engagement, and, therefore, momentum and adoption''\cite{kuksenok2016influence}. Therefore this work looks to expand the tradition of utilization of participatory design practises in the context of AI approaches to science by enabling a more flexible modeling approach while maintaining established key aspects of interpretability.

\section{Formative Study}

This section presents the findings of our inquiry into the analytic workflow of \pixl data by scientists focusing primarily on anomalies. 
We started this work with a series of contextual inquiry interviews \cite{wixon1990contextual} 
conducted in a cadence of approximately every two weeks over the course of
18 months to understand the different users of \pixl data. 
While we spoke to and collaborated with many dozens of scientists working with \pixl data, we focused our attention with five primary users: 
three spectroscopists who we will refer to as R1, R2, and R3;
a sedimentologist, R4; 
and a geochemist, R5.

Our research question entering into this study was to find what are the primary constraints that any modeling intervention in this context must satisfy in order to be useful within the scientists analytic workflow (RQ1). 
Through these interviews we were able to define three primary design constraints which elucidate the requirements for any anomaly detection method to be useful to \pixl scientists. While each design goal is firmly situated within the context of \pixl analysis, the underling rationale and aspects of \pixl data that inform each goal are not unique to \pixl and likely are applicable in a broad range of scientific applications. 

\subsection{Background on \pixl Science Workflow}
\begin{figure}
  \begin{center}
    \includegraphics[width=\columnwidth]{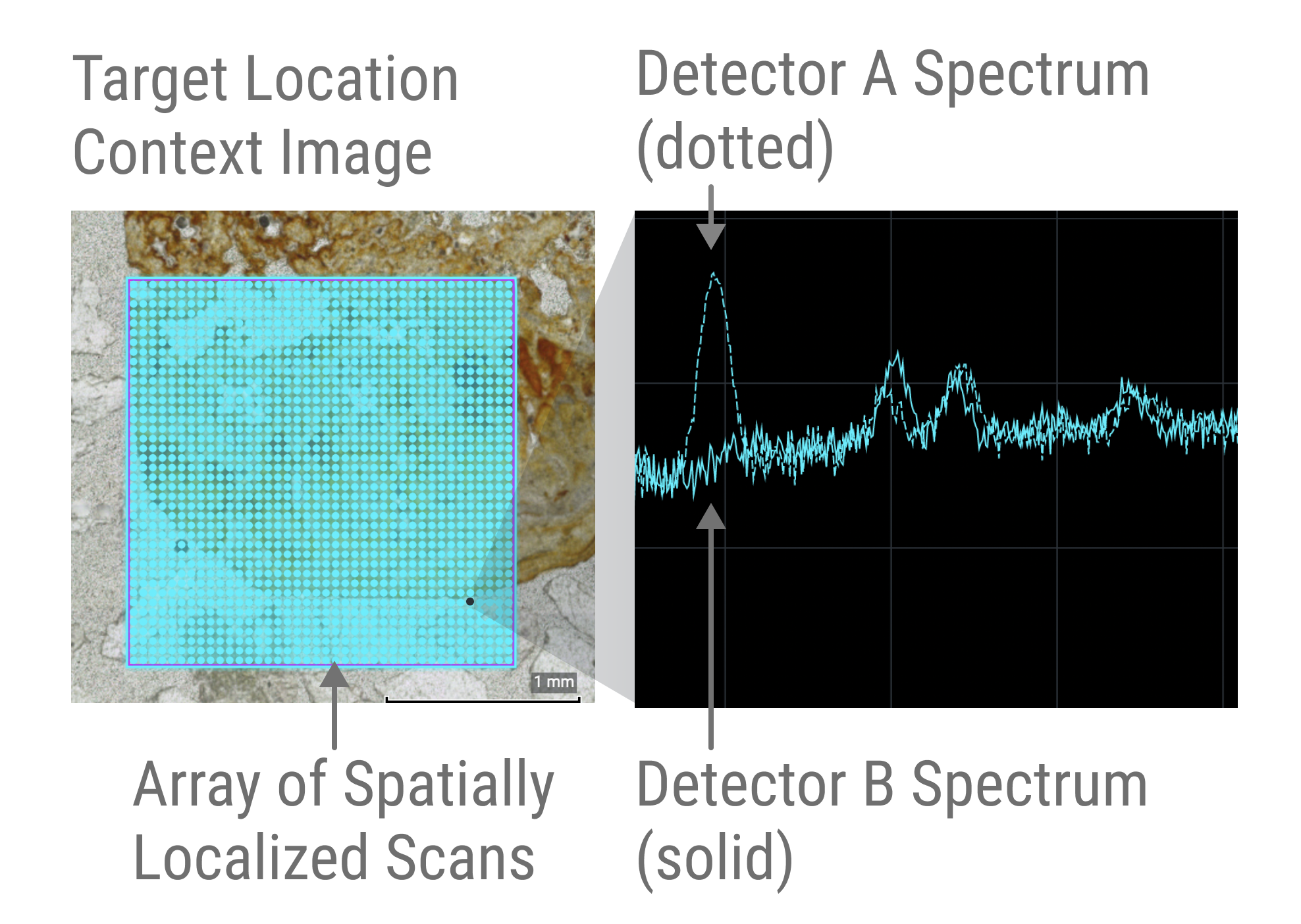}
  \end{center}
  \caption{Overview of data provided by \pixl instrument}
  \label{fig:data}
\end{figure}

In order to understand the context of the design goals for anomaly detection for \pixl, we must first establish some basic background of the data formats and processing steps scientists work with.
During an experiment, \pixl operates on a specially designated sample known as a \textit{target}. \pixl sends an X-ray beam into a location on the rock's surface. At each location, \pixl's X-ray beam causes the chemical elements in the target to fluoresce, which is captured in a data type called a \textit{spectrum} \cite{beckhoff2007handbook}, a 4096-index array of \textit{channels}. Each channel records the count of electrons, sensed at a distinct energy level measured in kilo electron-volts, or \textit{keV}. \pixl's \textit{A} and \textit{B} detectors each record the distinctive fluorescence patterns emitted by each point on a target from two different phase angles. These distinctive responses take the form of \textit{fluorescence peaks}, which are Gaussian peaks of a fixed width of channels over background measurements which are dependant on the chemical composition of the X-ray beam location. During an experiment, \pixl's camera also captures visible light images of a target. When data are returned to Earth, the X-ray beam geometry is reconstructed, and each spectral point is localized within each of the returned images. Overall, an experiment at a target returns a series of visible light images, and around 4,000 spectral points, each with an A and B spectra, and x and y coordinates within each image \cite{Allwood2020} (see Figure \ref{fig:data}).

Once an experiment is conducted with the \pixl instrument on Mars, the spectral and image data from that experiment is sent back to earth and analyzed primarily through the \pixlise analysis user interface \cite{PIXELATE}. This data is analyzed and transformed in multiple steps by different subject matter experts.
The first step in analysis, spectroscopists  translate the peaks in the spectra into elements that they believe are present in the target \cite{haschke2014laboratory}.

The spectroscopist then uses their list of elements to \textit{quantify} the spectra, translating the intensity of the various peaks into an empirical estimate of the percent of the total elemental mass each element constitutes in the target \cite{haschke2014laboratory}. \pixl uses the PIQUANT quantification algorithm \cite{elam2002piquant, heirwegh2021piquant}, and exposes those capabilities through the \pixlise user interface. 
Each spectra is quantified independently by PIQUANT, while  spectroscopists  determine the set of elements to quantify using the \textit{bulk sum} of all of the spectra in the dataset.

After the spectra have been quantified into elemental weight percents the broader science team begins to analyze the dataset. Since pure elements are uncommon in nature, the next task of the science team is to determine how the elements that have been detected and quantified combine to form some combination of the currently known 5700+ minerals \cite{Ruff2016}. The primary driver of this analysis is the quantified elemental weight percent map of the dataset, which is visualised using many standard Geology and Geochemistry data visualization techniques such as \emph{ternary diagrams} and \emph{heatmaps} combined with visualizations unique to \pixlise such as \emph{chord diagrams}. These quantitative signals are importantly augmented with additional signals from the color, shape, and texture of the rock in the images, or its \textit{morphology}, to consider mineral candidates. 

As candidate minerals emerge, scientists then designate them as \textit{regions of interest}, or ROI's, and evolve theories of the historical geologic processes that brought the rock and its individual ROIs into existence and altered them over time \cite{banfield2001:biosignatures}. These theories become increasingly dependent on the contextual and visual information and how minerals are situated as the discussion broadens to include Astrobiologists who can aggregate the details of the geochemistry, geophysics, and climate, to build a long-term theory of the broader context of the target, site and region and implications towards biological habitability.

\subsection{Design Goal 1: Focus on Raw Data Over Processed and Quantified Data (\labeltext{G1}{goal:raw_data})}
When considering the problem of anomaly detection in this workflow we first sought to understand what data structure to analyze, the raw spectra or the quantification. What we observed was that for many scientists the information from the \pixl instrument spectra was almost entirely mediated through elemental quantification of fluorescence phenomena and the visualization of these quantifications within \pixlise. This means that for the most part, anomalies are found by discovering unexpected results in a quantification and backtracking to find some non-fluorescent spectral phenomena that is causing an erroneous quantification. As R2 described:
\begin{quote}
    ``
    ... I'm getting quite a bit of activity in a barium oxide map, should I be getting that much... 
    ?
    ... So I might then 
    ...
    look at those individual spectra 
    ''
\end{quote}
 This method has obvious downsides, as any anomaly that causes an error in quantification not unusual enough to merit deeper scrutiny can propagate misleading information. Furthermore, the total amount of time and effort spent on downstream correction can be much greater than early detection, especially given the scale of data that \pixl produces. 
 Therefore we formed our first design constraint that we should \textbf{perform analysis on raw and unquantified spectra} in the hopes of catching phenomena that may be obfuscated through bulk sum quantification. 

\subsection{Design Goal 2: Robustness to Limited Ground Truth Labeling (\labeltext{G2}{goal:labels})}
The next constraint we found was that the actual amount of reliable ground truth labels currently existing in \pixl datasets is very limited. This is due to the requirement of manual processing by a small number of expert users to reach reliable conclusions, paired with the novel scale of data being produced by \pixl. R1 expressed their desire to help winnow down potential anomalies before digging into deeper analysis:
\begin{quote}
    ``What we want 
    ...
    in the automation phase is reduce 
    ...
    one data set a day with 5000 spectra down to
    ...
    a few spectra a week. Because 
    ...
    this is a multi-day interrogation [for] one data set, and so 
    ...
    with all of the other science outputs 
    ...
    ...
    you then want to go flagged for looking at later."
\end{quote}
Thus we formed our second constraint to be that our method must be \textbf{robust to a small number of ground truth labels} and thus provide a reasonable number of flags for experts to be able to precisely analyze.

\subsection{Design Goal 3: Allow Differentiation of Anomalies by Scientific Causal Process (\labeltext{G3}{goal:interpretation})}

Another major constraint that was emphatically expressed to us throughout our initial interviews was the vital importance of understanding why given anomalies may be presented within the context of scientific models. For instance R1 described why spectra are looked at manually currently due to the huge number of different ways to model a spectra and thus the requirement of background knowledge:
\begin{quote}
    ``Fluorescence data is fitted manually for a reason. 
    ... 
    you could have something with ...
    15 lines from rare earth elements in the spectra. There's always some expert user who knows something about the sample fitting the data. 
    ...
    because the combinations are infinite, that it's not something 
    ...
    automatically done.''
\end{quote}
 Furthermore, anomalous phenomena  may contain useful information in themselves about the physical processes that causes them, and thus may be worthy of analysis in their own right. We found that the non-fluorescent phenomena most often discovered when manually investigating quantification anomalies is diffraction, which is an effect often investigated on its own in the context of purpose built X-ray diffraction instruments \cite{bish2013x}, but which has a signal response sometimes overlapping fluorescence peaks. Thus we hypothesized that such spectral anomalies, if sufficiently differentiated, could be used as another source of auxiliary information similar to the visual context imaging within the mineral identification process. What is currently lacking is a way to find and characterize anomalies early in the pipeline and then utilize these detections for improved downstream analysis. R1 stated the goal similarly:
 \addtolength\leftmargini{-0.2in}
 \begin{quote}
     ``The ultimate test would be if fluorescence yields an ambiguous mineral that the diffraction can make unambiguous.''
 \end{quote}
 Therefore we determined it is very important for scientists to be able to not only find anomalies but \textbf{interpret and differentiate different kinds of anomalies} in the context of helping them understand the underlying science.

 \subsection{Comparison to the Existing Approach: Evaluating Standard Machine Learning Based Anomaly Detection Methods Using Design Goals}
 
 Once we had understood the domain and relevant design goals we sought to evaluate standard approaches to the problem of anomaly detection and take into account any potential issues (RQ2), and to guide the specific problems we need to address in out solution (RQ3). These standard approaches for anomaly detection and machine learning can be broadly broken down into traditional methods and deep learning based methods. Traditional methods tend to be designed for tabular data and are thus generally not well suited for analysis of raw data as required by design goal \ref{goal:raw_data}, while deep learning models are well known for their adaptability to complex non-tabular data formats. Deep learning based anomaly detection methods tend to follow the general structure of training a model to encode data into a compact representation to find structural patterns and outliers \cite{deepAnomalySurvey}.
The main classes of these methods that do not need extensive labeling (as required by design goal \ref{goal:labels}) are feature extraction methods and normality representation  methods\cite{deepAnomalySurvey}, both of which contain explicit assumptions that conflict with our design goals. 
 Feature extraction methods assume that ``The feature representations extracted by deep learning models preserve the discriminative information that helps separate anomalies from normal instances.'' This violates our finding from goal \ref{goal:labels}, where measurement is expensive and thus sufficient data to form a comprehensive feature set that includes rare and nuanced anomaly classes without explicit labels is not available. 
 Normality representing models take the form of latent space encoders such as auto-encoders or generative adversarial networks and assume that non-anomalous instances can be better represented and reconstructed by these encoding models than anomalous instances. We found this to not be true in practice, when considering the constraint of goal \ref{goal:labels} we had no way to preemptively sort out normal as opposed to anomalous instances for semi-supervision meaning anomalous instances must be included in the training data. This can cause problems as the model then will learn to represent those instances just as well as normal instances. This makes further sense when considering that the choice of reconstruction loss function is ``designed for dimension reduction or data
compression, rather than anomaly detection. As a result, the resulting representations are a generic
summarization of underlying regularities, which are not optimized for detecting irregularities.'' This property, when paired with the lack of supervised labels from \ref{goal:labels}, fundamentally violates goal \ref{goal:interpretation} as general purpose patterns explicitly do not prioritize truly rare or categorically different anomalies but will always prioritize either single point anomalies or simple sub-samples of classes of normal data which happen to be more rare.

Our goal of anomaly differentiation by scientific interpretation \ref{goal:interpretation} shows clearly how scientific end users care most about the \textit{underlying causal processes} as opposed to the most clear surface level empirical patterns. Lacking robust labeling, unsupervised methods all ultimately can do nothing but optimize in different ways for surface level empirical similarity without considering different causal processes. Thus even when such methods are able to discover some of the anomalies that are present, they fundamentally keep the task of sorting through the important vs unimportant classes of anomaly as a manual process. We hope to be able to reverse this order of operations, and allow scientists to differentiate the kinds of anomalies they care about first, and then detect them directly allowing for pre-sorted interpretations when processing new data.

 \section{Method}

When considering the weaknesses of deep learning and traditional data science based anomaly detection, a root cause of issues is that the problem framing involves either no direct input from scientists (which inevitably violates design goal \ref{goal:interpretation}) or only has input mediated through labeling (which in order to communicate required nuance would require a scale that violates design goal \ref{goal:labels}). 
Thus we set out to develop an alternative model development framework which can more effectively and efficiently incorporate scientists' prior knowledge into anomaly detection models (RQ3). 

A key insight in developing this framework is the differentiation between \textit{phenomena} and \textit{data}. 
Most existing methods focus purely on the space of data, and thus anomalies must be defined as individual data points. However scientists work in an ontology that is more abstracted from data space, where there are many underlying processes that can occur in the physical world and these processes can be measured in many different, or incomplete manners. What is important is not tied to a single datum, but rather what any subset or superset of data can imply about the underlying phenomena. Thus what we consider as phenomena can occur at multiple levels of scale. A single data point of a high dimension or complexity may contain within it multiple instances of different phenomena. 

This form of analysis is the one side of a trade-off. Data space analysis optimizes for \textit{completeness} by encoding all measurable information contained samples from dataset up to the limits of the size of the dataset without regard for \textit{correctness} or why any given datum has a particular set of features. By modeling phenomena we inherently limit completeness as only phenomena considered explicitly can be modeled, which must by necessity be less than the innumerable number of underlying factors that could be modeled given perfect knowledge. However what we gain is correctness, where phenomena that are of interest can be modeled more fittingly to their natural scale and be separated for interpretation by default rather than through post-hoc analysis of data space latent encodings.

\begin{figure*}
     \centering
     \includegraphics[width=1.0\textwidth]{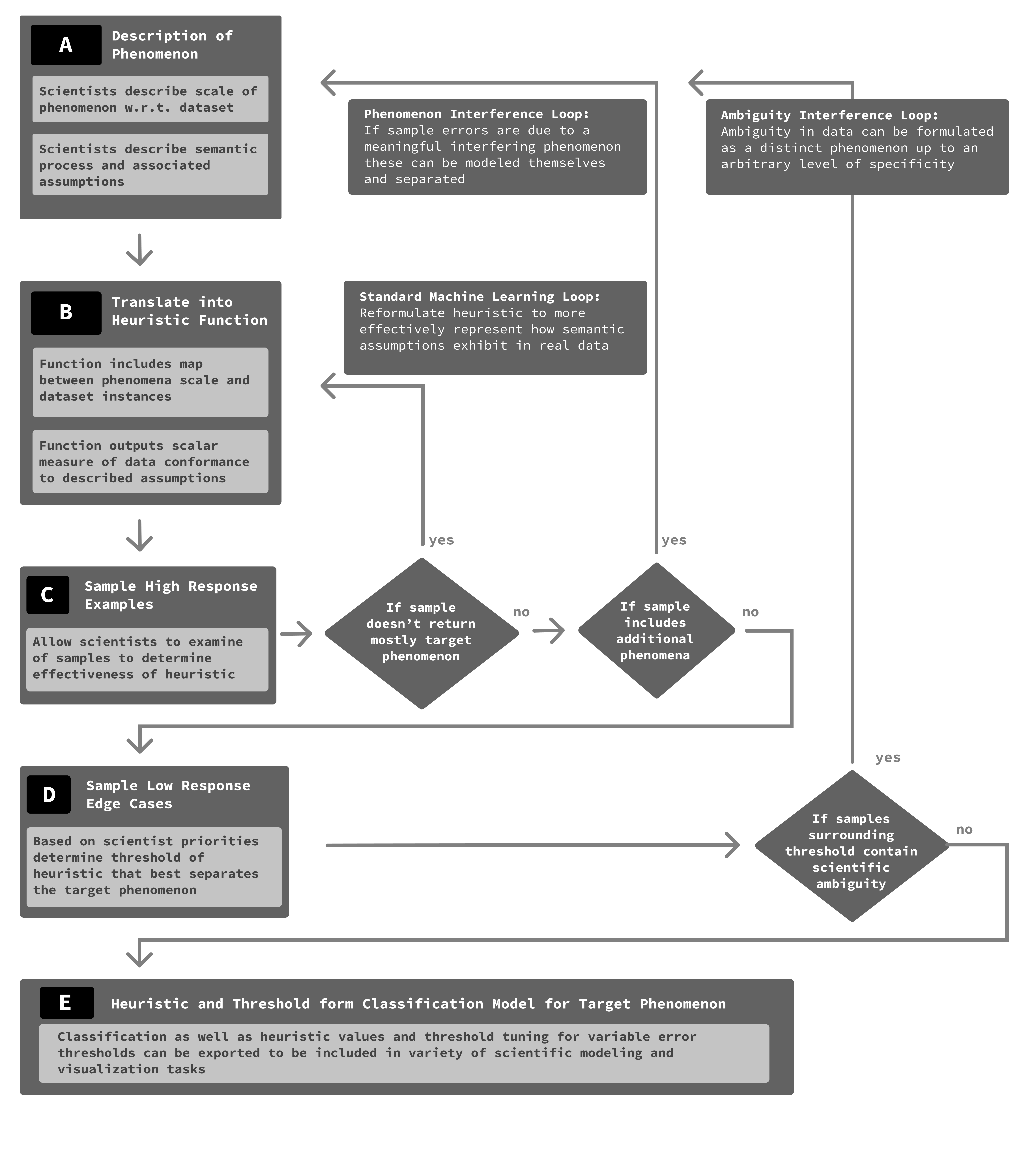}
     \caption{Overview of the ISHMAP Design Framework. Starting with scientist lead descriptions of phenomena (A), translated by developers into a computable heuristic function (B), which enables sampling of archetypal data instances (C), iterated until heuristics can provide a clear signal for the high response samples, followed by sampling of low response samples to determine a classification threshold (D), which when determined allows the return of a finalized model of the target phenomenon (E). An example of such a model as a result of this process can be seen in figure \ref{fig:pixl_diffraction_model}.}
     \label{fig:ishmap}
 \end{figure*}

Here we present a design framework, \textit{Iterative Semantic Heuristic Modeling of Anomalous Phenomena} (ISHMAP, see Figure \ref{fig:ishmap}), that can provide a template for developer and scientist collaboration to perform phenomena based anomaly analysis. Guided by the design goals laid out previously it can produce heuristic raw data feature extractors based on scientifically determined meaningful anomalous phenomena, and iterate adaptively based on the limited amount of scientist time available. We utilized this framework within the context of \pixl science and here discuss both the details of its was application within \pixl science to enable novel scientific discoveries as well as laying out the general principles as a potential resource for other collaborations guided by similar design goals. 

 \subsection{Scientist Description of Anomalous Phenomena}
  \label{sec:phenomena_formulation}
 The entry point into ISHMAP and the key differentiation between phenomena as opposed to data centered analysis is starting with scientist driven explicit identification of a specific anomalous phenomenon (Fig. \ref{fig:ishmap}A). The process begins by outlining a semantic class of anomalous phenomena within the ontology of the given scientific domain. This specific class is chosen by scientists as conveying some important information. 
 With \pixl we started with the phenomena of diffraction. Scientists chose to start with this phenomenon because diffraction only occurs with a particular and theoretically well-understood set of conditions involving crystal structure. This means that the presence of diffraction can tell scientists very important information about the physical structure of a sample that elemental composition alone cannot differentiate. 
 
 Once a phenomenon is decided upon the first characteristic that must be determined is the scale at which the phenomena is measured by the available data. Does the phenomenon occur as a specific kind of data point? Does is manifest as a pattern between adjacent points? Or does it occur, potentially multiple times, as a subset within a single data point? The scale of analysis is determined based on the prior understanding both of the physical phenomenon as well as the characteristics of the measurement methods producing the available data. This scale determination is absolutely essential to enabling properly interpretable heuristics, as it determines the \textit{input} of the heuristic function and thus the way it can express a phenomenon as the most natural explanation for how a phenomenon exhibits itself in a larger dataset. What must be decided for a scale determination is a \textbf{sampling procedure to extract from the primal dataset all potential instances of the target phenomenon}, as well as a map back to the primal dataset determining which parts of specific data point or points are being included in a sample. 
 For our example scientists know that diffraction occurs as distinct peaks within spectra, and based on the known resolution properties of the \pixl instrument, these peak responses are assumed to be discrete signals within a window of size 0.2 KeV which is the full-width-half-max size of detectable gaussian peaks for the detector \cite{7119099}. This means that the sampled input for a heuristic will be all contiguous windows of width 0.2 KeV which can be individually evaluated as potential diffraction peaks.
 
Once the correct scale of analysis is determined, scientists should then describe the differential causal process of the phenomenon with respect to the data measurement. What this entails is describing how the given phenomenon interacts with the measurement process and thus the differential between the described anomalous and non-anomalous data with respect to their underlying known or hypothesized causality. This description does not need to be extremely thorough, as it will later be translated through various lossy processes, it merely has to describe they primary ways in which this phenomenon differs from the default assumptions of the model, and how these differences manifest in the data. A well chosen scale determination will tend to greatly decrease the complexity of such descriptions when compared to descriptions that must work purely in the primal dataset. This description will form the basic starting point from which heuristics can be designed and iterated upon. 
For diffraction the causal process is well understood, diffraction is an effect that occurs when the \pixl instrument is particularly aligned with a crystal structure in the sample and \pixl sends X-rays of the correct frequency that resonates with the lattice the response will scatter with constructive interference forming a response peak at that resonant frequency. These response peaks are similar in shape to florescence as gaussian peaks with width determined by the detector resolution, however their causal process differs such that they can occur at arbitrary frequencies as opposed to solely at elemental florescent frequencies and the spatial dependence of the effect is sensitive enough that a diffraction response in one of \pixl's two detectors is very unlikely to occur at the same frequency in the other detector. Once scientists have formed such a description of the scale of a phenomenon and the causal forward process that generates differential data measurement, developers can proceed to the next step of ISHMAP (Fig. \ref{fig:ishmap}B).

 \subsection{Translation into Heuristic Model}
  \label{sec:heuristic_translation}

 After a definition of the phenomenon is provided by the scientists, it is then the job of the developer to translate this definition into a computable heuristic model (Fig. \ref{fig:ishmap}B). 
 This stage of the framework can take many different forms depending on the nature of the description provided. The only requirement being that the end result of an iteration of development be a program that takes as input a sample of data of the form set forth by the phenomenon scale characterization and output a scalar value proportional to how well a given sample conforms to the forward process of the anomaly characterization. The form of this program can depend on a number of factors including the format of forward process description, the amount and format of available data in the phenomenon scale, and compute resources available. When designing the initial heuristic for diffraction we chose to manually implement a statistical test of the assumptions provided in the previous step. We defined a function where given a window of spectrum counts of width 0.2 KeV, we test the hypothesis that one of the two \pixl detectors contains a statistically significant response peak above the spectrum's noise threshold while the other detector does not. This is done using a paired difference t-test \cite{student1908probable} between the two detectors pairwise over each channel in the window. This is is used as the counts in each channel are not independent since the underlying count for each channel is dependent on the X-ray frequency of that channel. We calculate and return the absolute value (since it does not matter which of the two detectors is the one where the diffraction is detected) of the t-statistic for the window as the measure of the statistical effect of potential diffraction. 

 While in the \pixl case the heuristic model took the form of hypothesis testing based on the assumptions provided by the scientists, there are many possible methods of formulating the heuristic. If the given scientific description is more easily expressed as Baysean priors then models (including deep learning based models) that utilize such probabilistic formulation may be useful. Alternatively if direct simulation techniques for particular anomalies are provided then direct similarity comparisons based on forward process simulation may be a better fit. The role of the heuristic model in the ISHMAP framework is not to enforce a single modeling paradigm for all phenomena, but rather to provide an interface for different models to work in ensemble within the larger framework. The important components are that models must be chosen to have the most appropriate phenomenon-scale inputs, have comparable scalar heuristic outputs, and provide some way of parameterization based on scientist priors with possibility of fine tuning. If all of these requirements are met then the heuristic can be calculated for each phenomenon-scale datum in the dataset and these pairs of data and heuristic value can be utilized in the next step (Fig. \ref{fig:ishmap}C).

 \subsection{Heuristic Model Evaluation from Sampling High Response Archetypes}
 \label{sec:high_response_eval}
 Given a version of the heuristic model the next step in ISHMAP is to evaluate the heuristic and subsequently determine the kind of iteration for model refinement (Fig. \ref{fig:ishmap}C). This is done first by sampling data from the high end of the distribution of heuristic responses. If the heuristic model is performing well these high response samples should form strong archetypes of the phenomena to be modeled. In this phase of the process scientists should be given an opportunity to inspect the class of model archetypes and determine the coherence of the class. If the high response samples are consistently determined to be good examples of the desired phenomenon then the heuristic model can be confirmed and moved on to the next threshold tuning phase. Otherwise the set of high response samples will contain instances of phenomena other than the target. 
 
 In this case scientists must then determine what else is being included. If the set contains a meaningful amount of instances of a distinct anomalous phenomenon then we can recursively iterate the ISHMAP procedure to model this other class and thus form a differentiation. This is exactly what occurred when evaluating the initial diffraction heuristic. During this phase R2 pointed out:
\begin{quote}
    ``
    And, okay, yeah, this is something actually that I think is good for me to point out to you, because the algorithm identifies this as a diffraction. And this is not diffraction. This is actually, I would say, an intensity mismatch that we're seeing these not just in that peak. But in some other locations, other peaks in the spectra, I think there's a little bit of intensity mismatch has to do with measuring the rough surface, like measuring like these larger grains.
    ''
\end{quote}
What we had found was an additional class of anomalous phenomena due to \textit{surface roughness}. This phenomenon was then subsequently modeled using ISHMAP, where the scientists' background knowledge informed us that surface roughness effects are frequency independent and thus the roughness phenomenon-space included whole spectra, and that the effect is expected to be a constant attenuation of the signal in a single detector. This background informed the heuristic roughness detector of calculating the mean detector difference across the whole frequency range of a spectrum, being essentially the maximum likelihood estimate of an assumed constant attenuation factor. 
This heuristic proved to be highly effective in the first iteration at distinguishing roughness effects. Upon completion of the deeper level of ISHMAP iteration with an effective roughness detector heuristic we could then separate out diffraction from roughness effects in the high response region leaving a now coherent class of diffraction instances using the original diffraction heuristic. Note that the phenomenon scale for roughness is not the same as that for diffraction, an example of how two different phenomena can have overlapping effects at one scale but can be easily differentiated at another scale. The recursive iteration and phenomenon centric structure of ISHMAP ensures that all heuristics model phenomena at their native scale while still being able to provide information between scales.

After differentiating all distinct classes of anomalies present in the high response sample set the primary heuristic model can be directly iterated to optimize differentiation with non-anomalous normal data and associated background noise. This iteration loop is flexible to the amount of scientist labeling available, as the baseline modeling assumptions assure a certain minimum semantic coherence of the initial heuristic making additional optimization optional while the flexibility of the heuristic modeling interface allow for models that can benefit from additional feedback when available. 
Thus the expected availability of downstream scientist labeling is an important constraint to consider when choosing a heuristic model class. 
This phase of ISHMAP is considered completed and we can continue to the next phase (Fig. \ref{fig:ishmap}D) after sufficient iteration to ensure the the high response samples of the heuristic model consistently represent archetypal instances of the target phenomenon.

\subsection{Heuristic Threshold Tuning from Sampling Moderate Response Edge Cases}
 \label{sec:threshold_tuning}

Once the heuristic model has been determined to be responding to the correct features there remains the task of determining a threshold for the heuristic response value for the purpose of categorization. This process is very similar to the previous phase where samples with heuristic values around a potential threshold are generated and then evaluated by scientists. If the samples remain highly coherent then the threshold can be lowered, while if there are insufficient numbers of the correct anomalies in a set of samples the threshold can be raised. This process can be iterated and tuned depending on the comparative importance placed by scientists to false positives and false negatives. This phase (Fig. \ref{fig:ishmap}D) still contains the same opportunity for iteration present in the high response phase (Fig. \ref{fig:ishmap}C), where if additional phenomena are found in the boundary region they can be recursively modeled providing cleaner edge case regions. Furthermore in this phase we additionally consider a special class of phenomenon we call ambiguities. In the region of a decision boundary for phenomena detection, scientists will often find instances which are ambiguous in certain respects making an underlying label difficult or impossible to assign. These ambiguous instances should not be confused with well defined samples whose heuristic values happen to be near the decision boundary.  Rather, ambiguities refer to examples where even the scientific ground truth may not be extremely clear. These phenomena introduce challenges for any classification scheme. In ISHMAP they are addressed by simply modeling ambiguity as a distinct phenomenon, where even if scientific priors will be by definition less robust, the features that scientists use to justify differences in interpretation are used as the basis for the heuristic modeling. Such features must necessarily exist as otherwise scientists would not have any empirical basis for uncertainty. This additional iterative phase is repeated until either no more ambiguous instances are found in which case a threshold can be determined exactly based on error tolerances, or is limited by scientist availability as often the ambiguous class may contain infinitely deep amount of different and idiosyncratic features that could be modeled. 

When sampling edge cases of diffraction scientists found two different ambiguous phenomena. The first were cases of single detector response where the potentially diffracting window was had a significant difference in detectors, but the difference was not strongly peaked as the prior expectation would be for a diffraction response and was instead more flat. This was then modeled by measuring the relative height of the detector difference peak in the center of the window to differentiate strongly vs weakly peaking instances. The second class of ambiguous phenomena were cases where instead of a consistent background in the non-diffracting detector, the baseline detector (detector with lower average count over the window) contained a small peak as well but attenuated compared to the other detector. This introduced different interpretations by different scientists and thus was modeled as ambiguous using a heuristic of looking at the coefficient of variation of the baseline detector in a window. After differentiating these primary classes of ambiguous phenomena the boundary region of the diffraction heuristic threshold became sufficiently clean to determine a threshold to classify diffraction, in this case with a bias placed on reducing false positive determinations. 

\begin{figure}
    \centering
    \includegraphics[width=0.95\columnwidth]{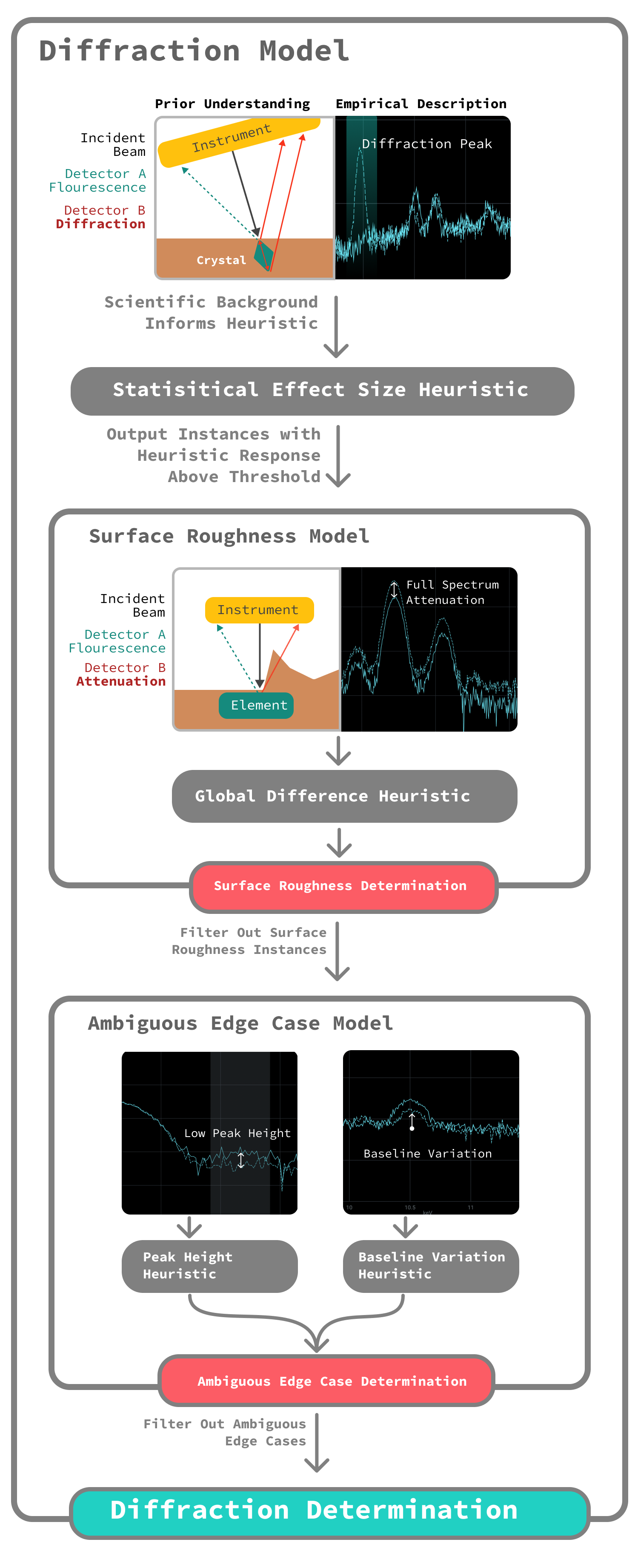}
    \caption{The architecture of the output model (Fig. \ref{fig:ishmap}E) of utilizing ISHMAP to detect diffraction peaks.}
    \label{fig:pixl_diffraction_model}
\end{figure}

Once all interfering phenomena have been modeled and differentiated, and a classification threshold is determined, the final output of ISHMAP is a deployable classification model (Fig. \ref{fig:ishmap}E). This model has the same fundamental classification structure as a decision tree, shown in figure \ref{fig:pixl_diffraction_model}, since the aforementioned phenomena differentiation is powered by similar recursive iterations of ISHMAP returned classifiers.

\section{Evaluation}

In order to evaluate our approach we both conducted a standard quantitative error analysis for the model component of the work (Section \ref{sec:quant_eval}), as well as a qualatative evaluation of the success of the final deployed system (Section \ref{sec:deployed_eval}). In doing so we do not aim to show that ISHMAP as a design framework is verifiably superior to other design approaches (as this would require counterfactual information of the success of many different design frameworks on this same problem which is well outside of the scope of this work). Instead we merely wish to show how ISHMAP \textit{is capable} of producing strong scientific outcomes in \textbf{this} application, and thus \textit{may} provide useful in further applications.

\subsection{Evaluation of Anomaly Detection Model}\label{sec:quant_eval}

After completing the ISHMAP procedure for the target phenomenon of diffraction, we are left with a classification model (Fig. \ref{fig:ishmap}E / Fig. \ref{fig:pixl_diffraction_model}).
In order to evaluate the real world accuracy of the model within the context of its use in mineral identification, and to avoid information leakage and test our model's generalizability, after developing our diffraction model using input from R2 and R3, we set to test the model using labels from additional scientists (R1, R4, and R5) on different datasets. Multiple different scientists were used in order to ensure reliable labels.
We presented the three different scientists a representative random sample of 213 spectra that were not used in the training process. The sample was balanced to include 107 spectra  uniformly randomly sampled among spectra determined by the model as containing diffraction peaks, and 106 sampled uniformly from spectra determined as not containing diffraction (through either exhibiting Surface Roughness, Ambiguity, or no anomaly). The scientists were then left to determine if any of the presented spectra contained diffraction. 
While all three scientists were presented with all 213 spectra, some spent considerably less time providing only a few labels on the presented spectra due to their time constraints. Furthermore we also found that the individual scientists had varying sensitivities for positively determining diffraction in cases of uncertainty and thus would often disagree on their determinations. Therefore we could not form reliable ground truth labels for 16 spectra which had only two labels from different scientists who disagreed, as well as for 45 spectra with only a single label from an individual scientist (which as we found is an unreliable indicator without a second opinion). This left 152 spectra (of which 144 had a total consensus and 8 had a majority determination) which we were able to 
use as a basis for evaluation. Of this reliable ground truth set, the model correctly predicted the presence or absence of diffraction with \textbf{93.4\%} accuracy.

These results match the qualitative experience that scientists expressed when examining the outputs of the model, as in an interview with R2 they expressed that:
\begin{quote}
    ``Your tool works very well, and is finding [diffraction peaks] in almost all cases, 
    ...
    And so there wasn't really anything that was being identified incorrectly.''
\end{quote}
This qualitative reliability and satisfaction from the perspective of scientist domain expert end users forms the most relevant evaluation of the efficacy of the tool when comparing to previous methods which do not present any systematic alternative baseline, and thus a strong example of the effectiveness of the ISHMAP Framework for designing useful models for scientific users.

\subsection{Impact of Deployed Interface}\label{sec:deployed_eval}

An important benefit of the ISHMAP collaborative design process is that once a model detecting a particular anomalous phenomena is complete, there is already assurance that the model is answering an actual problem of interest for scientists and collaborating scientists have a built in degree of ownership and buy-in to the technique\cite{sanders2008co}. This means that practical deployments of tools that utilize such a model are much more likely to result in adoption into the scientific workflow. To showcase how this collaborative design and scientific interpretation-first modeling approach can not only perform well with regards to general classification benchmarks but additionally result in meaningful new capabilities for scientific end users we can consider the deployment of our ISHMAP diffraction model within the \pixl science workflow. After finalizing the model we integrated the model outputs with the existing primary visual analytics toolkit used by \pixl science: \pixlise \cite{pixliseCOSPAR2021,ye2021pixlise-C,pixlisewww}. 

The first full version of the model and associated visualizations was deployed in November 2021, with preliminary versions available to scientists as early as June 2021. 
At the time of writing, it is used daily by over 97 \nasa scientists and \nasa-affiliated scientists across the globe collaborating on the \pixl science mission. Within the \pixl science group discussion board the functionality of our tool was mentioned over 80 times in the context of working group discussions, with 39 instances highlighting diffraction informed geo-science interpretations that would not have otherwise been easily visible and 28 instances of corrected spectroscopy and quantification error detection.
Now, whenever new data is beamed down from Mars, our model allows scientists to instantaneously discover diffraction peaks that can inform downstream mineral identification. 

\begin{figure*}
    \centering
    \includegraphics[width=1.0\textwidth]{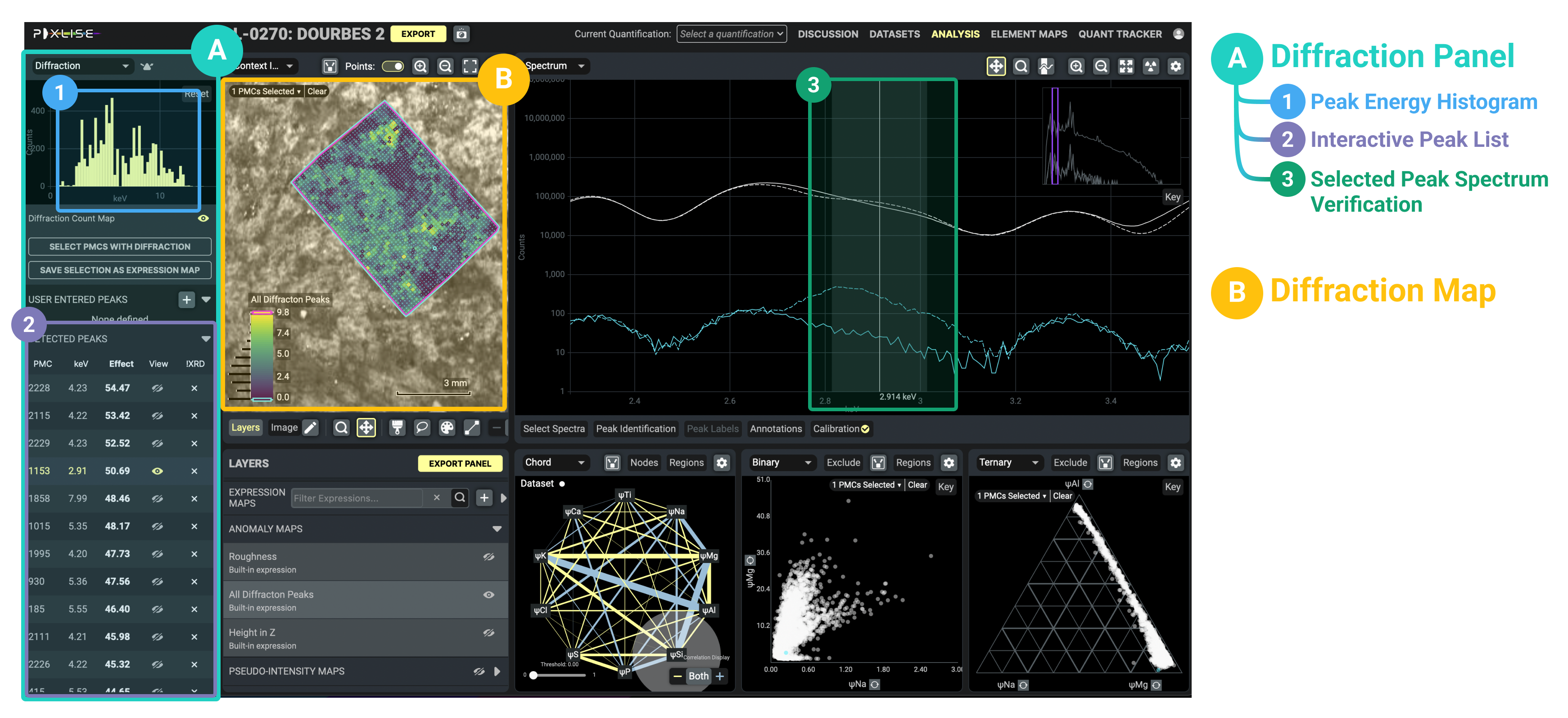}
    \caption{Overview of interface components within \pixlise Application displaying the Guillaumes Mars dataset. 
    (A) The \textbf{Diffraction Panel}  enables quick identification and verification of individual diffraction peaks or grouped similar peaks. 
    (B) The \textbf{Diffraction Map} displays the spatial distribution of diffraction peaks either over the whole spectrum range, for particular energy subsets, or in combination with custom defined expressions from other \pixlise analysis tools.}
    \label{fig:system_ui}
\end{figure*}

\subsection{Diffraction Panel Interface}
Immediately once a dataset is loaded into \pixlise, 
scientists can use
the \textit{diffraction panel} 
 (Fig.~\ref{fig:system_ui}A) to identify particular diffraction peaks. The diffraction panel includes a histogram of all of the energy levels where diffraction peaks have been detected.
The scientists can choose a set of energy ranges in the histogram to in turn select all locations which contain diffraction peaks in those ranges (Fig.~\ref{fig:system_ui}A.1). 
This allows scientists to quickly discover where different kinds of diffraction peaks, and thus minerals, are located (as the diffraction energy is a direct function of the crystal structure of the underlying mineral), as can be seen in figure \ref{fig:dourbes}. 
This is a particularly valuable feature for scientific interpretation enabled by the fact that our diffraction model works at the correct diffraction scale framing codified by ISHMAP as opposed to the default framing of anomalous spectra which a machine learning based model would utilize.
Scientists can then further verify individual peak identifications from a sortable list of the detected diffraction peaks (Fig.~\ref{fig:system_ui}A.2). 
Scientists can select a peak and then see its corresponding spectrum and energy location on the \pixlise spectrum view plot  (Fig.~\ref{fig:system_ui}A.3), 
and then confirm if this classification is a correct determination of diffraction or a false positive within the interface. 
This is required as even a very accurate model contains errors and allowing users to further refine the available data allows the model to be updated and improved continuously, as well as helping build trust with the scientists by ensuring that they always have the ability to override the interpretation of the model.

\subsection{Diffraction Map Visualization}
In addition to the peak-specific workflow enabled by the diffraction panel, 
we have also implemented a more high-level visualization of anomaly structure via the
\textit{diffraction map} interface
(Fig.~\ref{fig:system_ui}B). 
Within the \pixl science team, diffraction maps have become an accessible, shareable, and invaluable piece of information for the process of mineral identification as new data comes down from \pixl. 

The diffraction map visualization overlays a heatmap of the density of diffraction peaks or surface roughness anomalies at each beam location on top of the visual context image. 
This allows scientists to quickly find clusters of diffraction that are indicative of a crystal grain, group the locations within that cluster, and create regions of interest that can be applied towards further analysis. The diffraction map has become the preferred method of scientists to share findings of crystal structure, as R3 commented when looking at the diffraction map for the Beaujeu \cite{pdsBeaujeu} dataset:
\begin{quote}
    ``I found an interesting correlation between the regions marked with a high number of diffraction peaks, and the regions that we have geochemically identified as plagioclase... It is great to see the usefulness of the diffraction peak detection algorithm in practice.''
\end{quote}

\begin{figure*}
    \centering
    \includegraphics[width=1.0\textwidth]{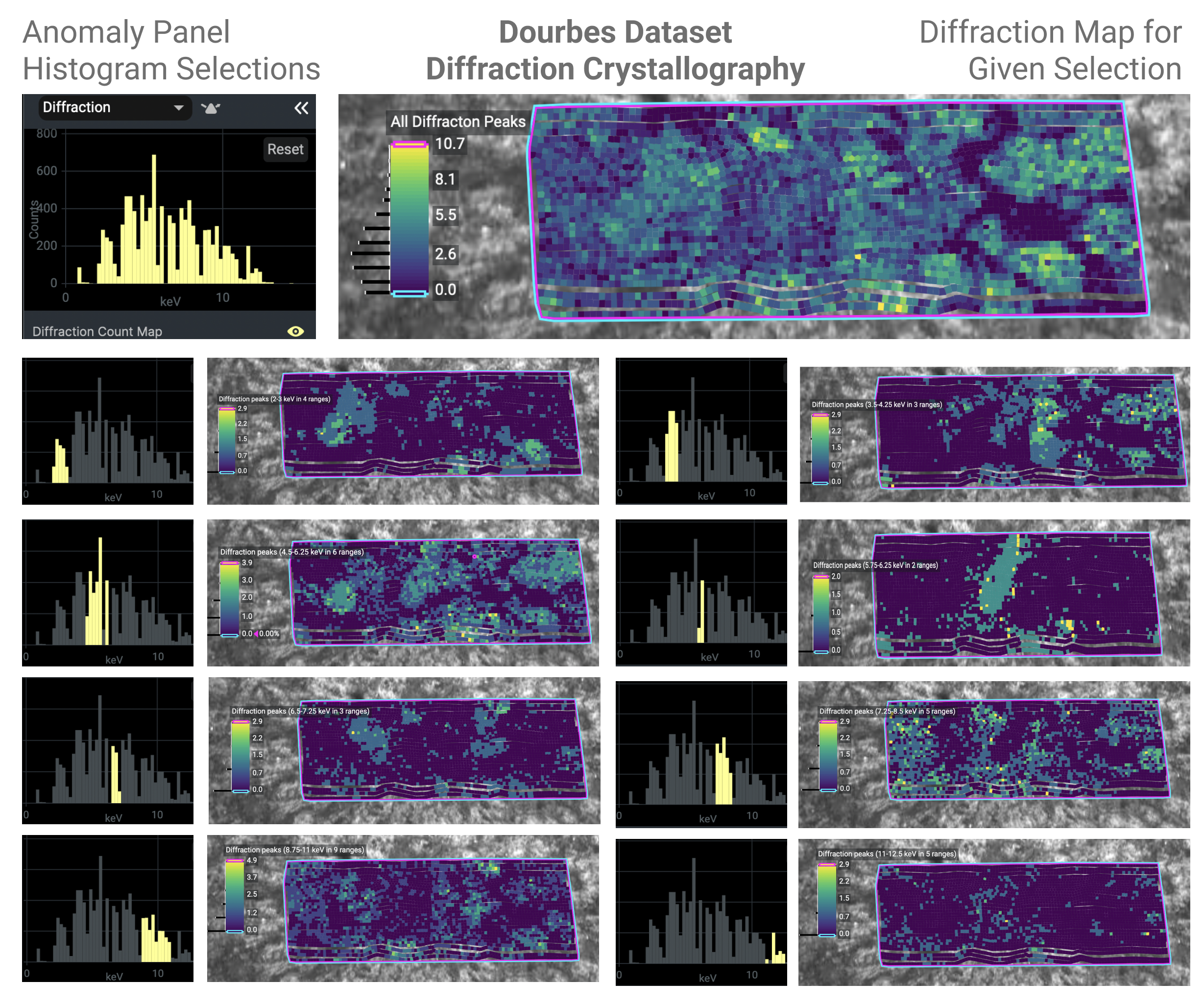}
    \caption{Screenshots of diffraction maps of the Dourbes\cite{nasaDourbes} dataset formed from different selections in the diffraction panel of diffraction peak energies. What can be see is rich information regarding the spatial distribution of diffraction peaks at different energy levels, with peaks in close energy clusters also clustering spatially. This implies the presence of unique crystal grains which can be readily seen using these diffraction maps.}
    \label{fig:dourbes}
\end{figure*}

These maps can be customized in a number of ways. 
The default view when starting is to show the density of diffraction present at all energy levels, this is the most broadly applicable when lacking a particularly strong prior about the specific crystallography of the target.
However if a scientist has a hypothesis about a particular crystal configuration which would predict diffraction at predictable frequencies there are two ways to visualise diffraction with more specificity. 
\pixlise contains a custom domain-specific language for custom maps of expressions. 
The output of the diffraction model is a supported query within this language and allows scientists to integrate anomaly information with other existing analysis.
Additionally, the diffraction panel histogram selection supports the creation of maps directly. 
By being able to see the distribution of diffraction peak energy, scientists can analyze the distribution and find clusters without a-priori knowledge, creating maps that rather than showing the overall crystallographic structure of the sample can isolate particulate grains (Fig.. \ref{fig:dourbes}). 

\subsection{Enabling Ongoing Scientific Discoveries about Martian Geology}

In November of 2021, the \pixl conducted a series of XRF scans of a sample with the codename \textbf{Dourbes} \cite{nasaDourbes} at the Séítah formation \cite{nasaSeitah} in the floor of the Jezero crater on Mars. This location presented an acute issue for the problem of mineral identification. 
Due to weathering, it is impossible to clearly identify crystal grains from the context imaging, and XRF information cannot sufficiently differentiate between all relevant physical properties. 
This information is extremely important to make inferences about the geological history of the site and has formed a significant challenge to previous Mars missions in similar situations. 

Fortunately, due to the additional data collected by the \pixl instrument and the development of our suite of tools scientists were rapidly able to visualize the crystal structure of the sample with the diffraction map (Fig. \ref{fig:dourbes}).
The diffraction map functionality enabled robust spatial comparison of diffraction with elemental analysis, and thus scientists could make strong claims about particular grains of elements, their crystal properties, and thus identified mineralogy. 
By going beyond the information available in standard fluorescence and elemental quantification, this comparison provided decisive evidence about the mineralogy of Séítah formation rocks, as expressed by a \pixl scientist in  
a recent abstract to be published at a top-tier scientific journal\cite{tice2022alteration}:
\begin{quote}
    ``Collocated crystal sizes and mineral identities are critical for interpreting textural relationships in
rocks and testing geological hypotheses, but it has been previously impossible to unambiguously
constrain these properties ... Here we demonstrate that
diffracted and fluoresced x-rays detected by the \pixl instrument ... provide information about the presence or absence of
coherent crystalline domains in various minerals.''
\end{quote}

This finding has formed the central component of continuing research within the \pixl science team and is functionally enabled by the diffraction detection and visualization capabilities powered by our model. The effectiveness of the model in integrating with existing scientific workflows and assisting in high impact analysis immediately upon deployment further provides strong evidence of the effectiveness of the ISHMAP design framework. 

\section{Discussion}

\subsection{Generalizability of Design Goals}
While this work has showcased a single specific successful application, we hope that there are a number of useful insights from our solution that that can be utilized in other applications as well.
In order to evaluate the applicability of our framework for other use cases the key deciding factors should be through the alignment of our outlined design goals. While we developed these goals strictly within the context of embedded user research for the particular domain of \pixl scientists, we justified the formulation of each goal based on aspects of scientists' workflows that we found to be common among the different individuals and specialities within the fairly diverse \pixl team as well as on aspects that, while exhibited in this specific workflow, are not necessarily unique to it. 
For instance our design goal \ref{goal:raw_data} is based on the observation that anomalies are more easily missed the further a dataset is from the `native' scale of the anomalous phenomenon. Since essentially by definition anomalies are classes of phenomena that default assumptions do not apply to, any processing steps are likely to introduce errors with respect to anomaly detection. We can then say that a focus on raw data is not just an important design consideration for \pixl, but for any analytic workflow that includes steps that processes data in an irreversible manner based on violable assumptions. 
Our design goal \ref{goal:labels} essentially formalizes an extremely common design constraint of limited data and label availability. The entire branch of unsupervised machine learning studies various implications for modeling with such a restriction. So while discovering that this constraint was applicable to our specific domain is extremely important, it is also clear that there are many other domains that share a similar requirement. 
Finally our design goal \ref{goal:interpretation} similarly expresses a well established and known weakness of deep learning based anomaly detection\cite{deepAnomalySurvey} and shows why it is important in our use specific case. 
What this all implies is that while the goals we developed are in one respect specifically tied to the \pixl science domain, we expect that there are very likely a large number of other domains who share these goals as well, and it is in these domains where ISHMAP may be a useful tool to structure model development.

\subsection{Limitations of ISHMAP}
While we present the substantial potential effectiveness of the ISHMAP framework, like all frameworks it is not universally applicable and has limitations to where is should be utilized. Of course the primary drivers of whether ISHMAP is appropriate is whether the design goals laid out are of relevance. In prioritizing these goals other potential priorities are de-emphasized. In particular the ISHAMP process can only help in discovering known anomaly classes, as well as anomaly classes that are discovered during the ISHMAP process. There are many domains, including scientific domains, where data sources are sufficiently novel to contain many potential `unknown unknown' classes of anomaly that users do not have a prior expectation for. Since ISHMAP relies on explicit description of known phenomena, such unknown anomalies cannot be captured reliably. In such applications it has been shown that more pure deep learning based methods can be highly effective in discovering such unique or point anomalies \cite{deepAnomalySurvey}. 

Furthermore the collaborative process may have substantial organizational overhead due to the requirement for consistent iteration and feedback between developers and scientists. Depending on the nature of a collaboration this overhead may occasionally present a greater manual effort burden than the effort of just providing more ground truth labels, undermining  design goal \ref{goal:labels}. Thus when making the choice of whether to undertake an ISHMAP collaboration both sides must consider both the technical and organizational nature of the problem to determine if it is the right fit.

\subsection{Opportunities for Future Work}
In presenting the ISHMAP framework we have only taken a first step in improving scientific anomaly detection with human-centered-AI methodology. While the results of our implemented detection tool with the \pixl team was a clear success, it only formed a proof-of-concept for the methodology. We encourage researchers to replicate, evaluate, and refine the methodology in additional domains. Additionally the framework itself presents clear opportunities for development. The flexibility of the framework makes it amenable to many different forms of utilization and deployment, and so future work to fine tune the most effective processes both technical and procedural for scientific prior encoding, heuristic generation, and sample evaluation may greatly assist in more efficient and effective utilization of ISHMAP. Indeed, while ISHMAP was developed in order to address the shortcomings of pure deep learning based approaches, studying how to integrate deep learning models and all their expressive power within this more interpretation-focused framework may allow for the best of both methods. Finally, for the current formulation of ISHMAP, while the expertise of scientists is absolutely essential, the role of the model designer/developer is comparatively procedural. This leaves a potential opportunity to develop automated tools or interactive interfaces for scientists to engage in their side of the ISHMAP procedure entirely independently, which would massively increase the potential for science teams to develop their own robust and interpretable anomaly detection models. 

\subsection{Reproducibility}

The code for discussed in this work 
is distributed across a number of repositories in the broader \pixlise project 
that is open-sourcing all of its constituent repositories at \url{https://github.com/pixlise}\cite{ryan_stonebraker_2023_7539750, tom_barber_2022_6959138}. 
Additionally the deployed \pixlise tool itself is publicly accessible. Anyone can request an account at \url{https://www.pixlise.org/} and use all of the functionality of PIXLISE, including the anomaly detection functionality discussed in this work, on any public datasets. Most of the datasets used as examples in this work are publicly available either on PIXLISE or in raw data form directly from the NASA Planetary Data System (PDS) at \url{https://pds-geosciences.wustl.edu/m2020/urn-nasa-pds-mars2020_pixl/}.

\section{Conclusion}

Many of history's most important scientific discoveries can be attributed to the thorough analysis of anomalies\cite{kuhn1970structure}. Today, as scientific datasets get larger and more complex, so too do the methods used to find the anomalies within them, sometimes at the expense of the ability to interpret and explain the anomalies\cite{deepAnomalySurvey} which is a fundamental component of scientific analysis. In this work we sought to integrate human-computer interaction methodologies with the state of the art in AI to develop a method for anomaly detection that is both effective and interpretable. 
In collaboration with a world leading science team at \nasa, we conducted extensive user research to understand their specific analytic workflow, and developing design goals that represent the needs of scientists with respect to interpretable anomaly detection. Based on these design goals we introduced ISHMAP, a novel design framework for the development of scientific anomaly detection models. By utilizing ISHMAP to develop an anomaly detection toolkit used daily by \nasa scientists around the world and contributing to ongoing scientific discoveries, we showcased a proof of concept for a method to enable better science by taking a human-centered approach to both the technical and scientific problems of anomaly detection which we hope can assist scientists and researchers looking to not only detect, but understand anomalies in their data.

\begin{acks}
The research was carried out in part at the Jet Propulsion Laboratory, California Institute of Technology, under a contract with the National Aeronautics and Space Administration (80NM0018D0004)
\end{acks}

\bibliographystyle{ACM-Reference-Format}
\bibliography{references}

\newpage



\end{document}